\documentclass[12pt]{iopart}

\usepackage{iopams}
\usepackage{longtable}
\usepackage[latin1]{inputenc}   % For accented characters
\usepackage{graphicx}  % For including figure files
\usepackage{dcolumn}  % For aligning table column on decimal point 
\usepackage{nicefrac}   % For nice small fractions
\graphicspath{{./}{figure/}}
\usepackage[dvips]{color} % Colored text for corrections
\definecolor{red}{rgb}{0.85,.1,0}
\definecolor{cblue}{named}{CadetBlue}
\usepackage{iopams}
\newcommand{\fetese}{Fe$_{1+y}$\-Se$_{x}$\-Te$_{1-x}$}
\renewcommand{\vec}[1]{\boldsymbol #1}

\begin{document}
\title{Magnetic glassy phase in {Fe$_{1+y}$\-Se$_{x}$\-Te$_{1-x}$} single crystals}
\author{G~Lamura$^1$, T~Shiroka$^{2,3}$, P~Bonf\`a$^4$, S~Sanna$^5$, F~Bernardini$^6$, R~De~Renzi$^4$, 
R~Viennois$^{7,8}$, E~Giannini$^7$, A~Piriou$^7$, N~Emery$^9$, M~R~Cimberle$^1$ and M~Putti$^1$}
\address{$^1$CNR-SPIN and Universit\`a di Genova, via Dodecaneso 33, I-16146 Genova, Italy}
\address{$^{2}$Laboratorium f\"ur Festk\"orperphysik, ETH-H\"onggerberg, CH-8093 Z\"urich, Switzerland}
\address{$^{3}$Paul Scherrer Institut, CH-5232 Villigen PSI, Switzerland}
\address{$^4$Dipartimento di Fisica and Unit\`a CNISM di Parma, I-43124 Parma, Italy}
\address{$^5$Dipartimento di Fisica and Unit\`a CNISM di Pavia, I-27100 Pavia, Italy}
\address{$^6$Dipartimento di Fisica, Universit\`a di Cagliari, I-09042 Monserrato (CA), Italy}
\address{$^{7}$D\'epartement de Physique de la Mati\`ere Condens\'ee, Universit\'e de Gen\`eve, 24 Quai Ernest-Ansermet, CH-1211 Geneva, Switzerland}
\address{$^{8}$Institut C.\ Gerhardt, UMR 5253, Universit\'e Montpellier II, place Eug\`ene Bataillon, 
F-34095 Montpellier, France}
\address{$^9$ICMPE/GESMAT, UMR 7182 CNRS-UPEC, 2-8 rue Henri Dunant, F-94320 Thiais, France}
\ead{tshiroka@phys.ethz.ch}
\begin{abstract}
The evolution of the magnetic order in {Fe$_{1+y}$\-Se$_{x}$\-Te$_{1-x}$} crystals as a function of Se content was investigated by means of ac/dc magnetometry and muon-spin spectroscopy. Experimental results and self-consistent DFT calculations both indicate that muons are implanted in vacant iron-excess sites, where they probe a local field mainly of dipolar origin, resulting from an antiferromagnetic (AFM) bicollinear arrangement of iron spins. This long-range AFM phase disorders progressively with increasing Se content. At the same time all the tested samples manifest a marked glassy character that vanishes for high Se contents. The presence of local electronic/compositional inhomogeneities most likely favours the growth of clusters whose magnetic moment ``freezes'' at low temperature. This glassy magnetic phase justifies both the coherent muon precession seen at short times in the asymmetry data, as well as the glassy behaviour evidenced by both dc and ac magnetometry.
\end{abstract}
\submitto{\JPCM}
\pacs{74.25.Dw, 74.25.Ha, 76.75.+i}
%74.25.Dw Superconductivity phase diagrams
%74.25.Ha Magnetic properties
%75.47.Lx Manganites. Pnictides
%75.50.-y Studies of specific magnetic materials
%76.90.+d Other topics in magnetic resonances and relaxations (restricted to new topics in section 76)
\maketitle

\section{\label{sec:intro}Introduction}
Among the recently discovered Fe-based superconductors, the iron chalcogenide family \fetese\ is structurally one of the simplest \cite{Johnston2010}, being characterized by layers of square lattice of Fe ions, tetrahedrally coordinated with a chalcogen. At room temperature its tetragonal cell belongs to the \textit{P}4/\textit{nmm} space group, with iron atoms occupying also an additional site in the Te plane \cite{Fruchart75}.
The occupation of this additional Fe site can be reduced by growing crystals starting from Fe-deficient and Se-rich compositions \cite{giannini2010}. Tiny variations in both excess iron ($y$) and Se substitution rate ($x$) strongly affect the structural, electronic and magnetic properties of \fetese.
Upon cooling,  the room-temperature tetragonal crystal structure of the parent compound ($x=0$) becomes either monoclinic (\textit{P}2/1\textit{m}), for $y \lesssim 0.125$ \cite{Fruchart75,PDai2009}, or orthorhombic (\textit{Cmme}), for $y > 0.125$ \cite{Bao2009,Martinelli2010}. This transition to a lower-symmetry structure is highly correlated with a magnetic order that involves 
the Fe moments below $\sim 70$~K. It consists in a bicollinear commensurate antiferromagnetic order, as predicted theoretically \cite{Ma2009} and proved experimentally by neutron diffraction \cite{PDai2009,Bao2009}. This magnetic order becomes incommensurate when the concentration of excess Fe atoms is larger than a critical value ($y \sim 0.1076$) \cite{PDai2009,Fang2009}. At the same time, a partial substitution of Te with Se progressively suppresses the tetragonal-to-monoclinic structural
transition, changes the nature of AFM order into an incommensurate short-range (SR) one and induces superconductivity at low temperatures. 
Several neutron scattering \cite{Martinelli2010,LumsdenNP, Liu2010,PDai2010,Tranquada2010,PDai2009,Bao2009,Tranquada2009,TranquadaJPSJ} and zero-field muon spectroscopy (ZF-$\mu$SR) \cite{Babkevich2010,Khasanov2010,Khasanov2009} studies delineate a picture in which the static long-range AFM order evolves into a short-range incommensurate one for  excess iron values about 0.1 and/or for Se substitution rates exceeding 0.1, with the magnetic phase eventually coexisting with superconductivity \cite{Khasanov2010,Khasanov2009}. This picture is made even more complex by the presence of a spin-glass phase, as suggested by dc magnetization data for both $x>0.15$ and $y>0.01$ \cite{TranquadaRPG,TranquadaJPSJ,PauloseJAPS,PauloseEPL}. Moreover, recent extended X-ray absorption fine-structure (EXAFS) measurements as a function of temperature have evidenced local structural inhomogeneities in crystallographically homogeneous \fetese\ samples: Se and Te occupy distinct sites with consequently inhomogeneous distribution of the Fe-Se/Te bonds \cite{Joseph2010,Iadecola2011}. Finally, scanning transmission electron microscopy (STEM) and electron energy-loss spectroscopy (EELS) on \fetese\ single crystals indicate the presence of nanometer-scale phase separation and chemical inhomogeneity \cite{Hu2011}. These intrinsic inhomogeneities most likely play a major role in determining the fundamental electronic structure and the magnetic properties of this compound.\\
To date the nature of the magnetic order, crucial for understanding the whole phase diagram, is not yet clear and is still under debate. To shed light on the evolution of the magnetic ground state as a function of the iron excess and Se doping, we carried out a systematic study by means of ac and dc magnetometry measurements as well as by zero- and transverse-field $\mu$SR spectroscopy on a series of \fetese\ single crystals.
Zero-field $\mu$SR represents a unique and powerful local probe for following the evolution of the magnetic order from long-ranged commensurate, through long-ranged incommensurate, down to short-range order, able to distinguish also the presence of possible spin-glass/cluster spin-glass phases \cite{UEMURA,CONERI,SannaCLUSTER}.
By combining this technique with complementary magnetization measurements, it was 
possible to detect the presence of a glassy magnetic state induced by Se substitution.
After describing the sample preparation and characterization in Sec.~\ref{ssec:preparation}, we focus on ac/dc magnetometry and $\mu$SR experimental results in the rest of  Sec.~\ref{sec:exp_details}. Successively, we discuss our main findings in Sec.~\ref{sec:discussion}, which are then summarized in the conclusions.
\begin{figure*}
\centering
\includegraphics[width=0.45\textwidth]{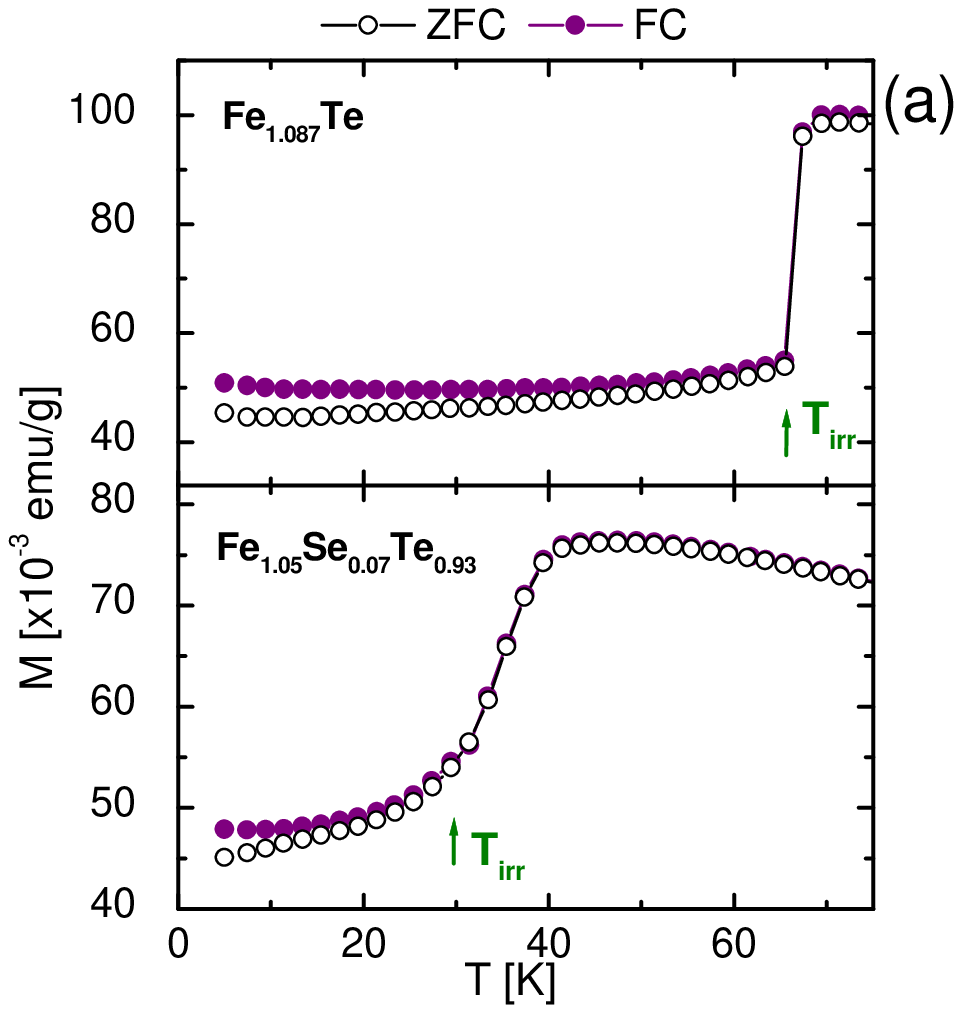} %DCmag_a
\includegraphics[width=0.45\textwidth]{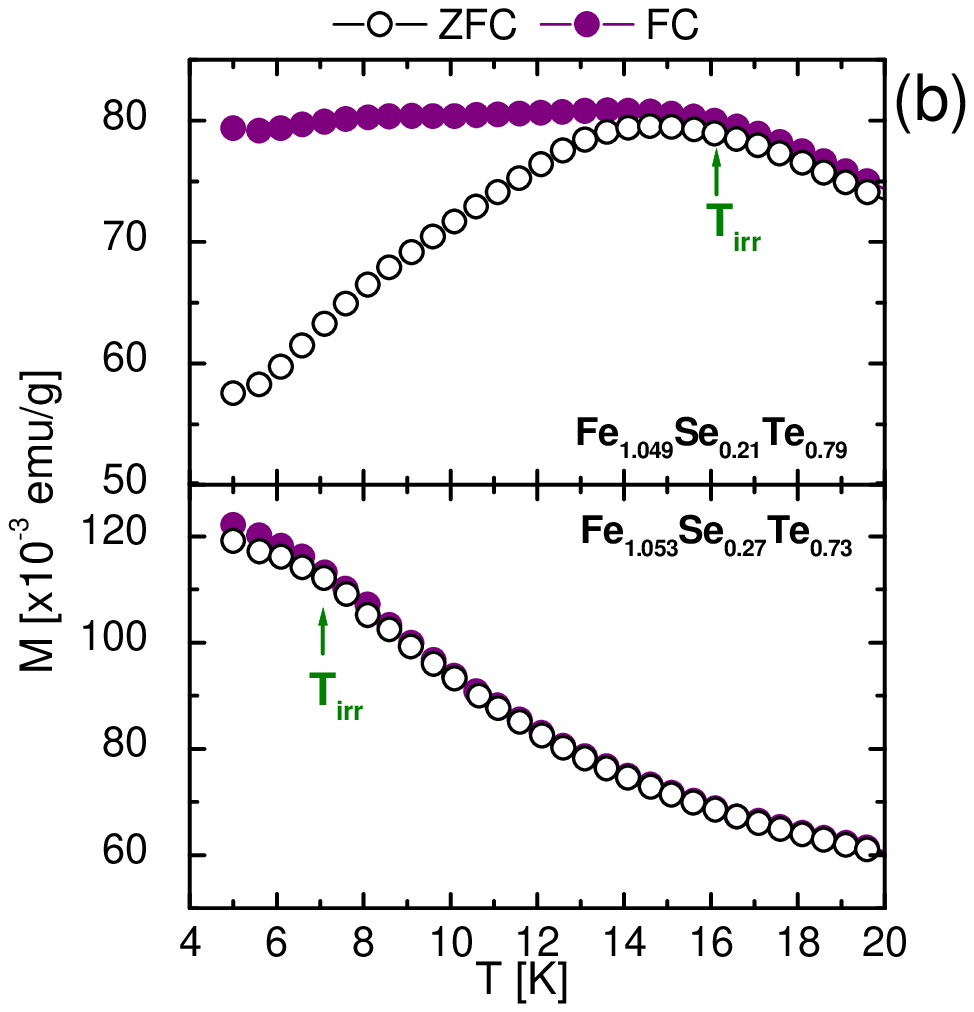} %DCmag_b
\caption{\label{fig:DCmagn}(Colour online) \fetese\ DC magnetization data for
$x=0$, 0.07 (panel a) and  0.21, 0.27 (panel b), respectively. Measurements were carried out in an external magnetic field of 200 mT. The vertical arrows indicate the irreversibility temperature $T_{\mathrm{irr}}$ (see text).}
\end{figure*}
\section{\label{sec:exp_details}Experimental details}
\subsection{\label{ssec:preparation}Sample preparation and characterization}
High-quality, single-phase \fetese\ crystals were grown starting from two different Fe:(Te, Se) ratios, 1:1 and 0.9:1, corresponding to two different compositions of the Te-rich flux. Crystalline samples were grown inside vacuum-sealed quartz crucibles in a vertical temperature gradient, by using a modified Bridgman method. The crucibles were slowly cooled from the liquid state ($T_{\mathrm{melt}} = 930$--960${}^{\circ}$C, depending on Se content) at cooling rates between 1 and 5${}^{\circ}$C/h. Further details on sample preparation are reported in Ref.~\cite{giannini2010}.
Crystals with actual compositions in the range $0.03\lesssim y  \lesssim 0.1$ and $0 \lesssim x  \lesssim 0.3$ could be obtained. Successively they were thoroughly characterized via structural and chemical analysis, as well as by transport and magnetic measurements. The actual composition was obtained from single-crystal X-ray diffraction pattern refinement, as described in Ref.~\cite{giannini2010}. Transport properties (resistivity and thermopower) have been reported in Ref.~\cite{Pallecchi2009}. Selected mm-size crystals of high purity and homogeneity were used for the macroscopic (magnetometry) and microscopic (muon-spin spectroscopy)
magnetic measurements reported in the present work.
\subsection{\label{ssec:DC_AC_su}Ac and dc magnetometry measurements}
\begin{figure}
\centering
\includegraphics[width=0.5\textwidth]{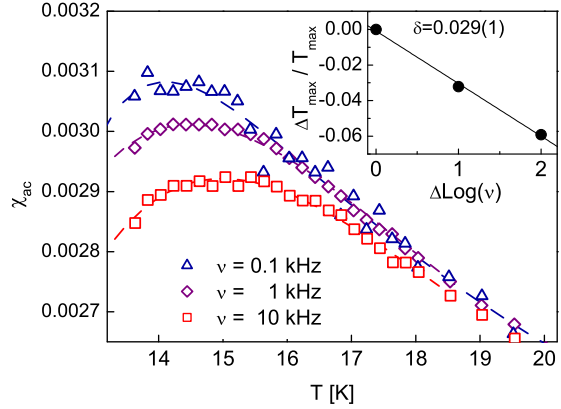} %chiAC
\caption{\label{fig:ACmagn}(Colour online) AC susceptibility of \fetese\ for $y \simeq 0.049$ and $x = 0.21$. Inset: frequency dependence of $T_{f}$, the temperature of $\chi_{\mathrm{ac}}$ maximum, as measured at $\nu=0.1$, 1, and 10~kHz (see Sec.~\ref{sec:discussion} for details).}
\end{figure}

The macroscopic magnetic properties of the \fetese\ samples were investigated via dc and ac magnetometry by means of a superconducting quantum interference device (SQUID) and a physical property measurement system (PPMS), respectively (both by Quantum Design).
In case of dc magnetization we adopted a parallel-field configuration, with the sample surface being parallel to the applied magnetic field. At low-temperature, low-field magnetization measurements indicate that all the Se-substituted samples are superconductors, whose transition temperature and shielding fraction increase with Se content, with the latter remaining far from the ideal $-1$ value (see table~\ref{tab:MagnProp} and \ref{ssec:DCmagnetization}). At higher applied fields all samples present interesting, albeit complicated, magnetic properties. In figure~\ref{fig:DCmagn} we show the zero-field cooled (ZFC) and the field cooled (FC) low-temperature bulk magnetic response at a static magnetic field of 0.2 T. Samples $x=0$ and 0.07 display signatures of an AFM order at about 70 and 40 K, respectively. Superimposed to this magnetic order, there is a hysteretic behaviour, as evinced from the bifurcation between the ZFC and FC dc magnetization curves. In conventional spin glasses, the temperature below which the ZFC and FC curves bifurcate is defined as the irreversibility temperature, $T_{\mathrm{irr}}$ \cite{MYDOSH}. Interestingly, in our case $T_{\mathrm{irr}}$ decreases continuously with increasing Se content. To better clarify this possible low-temperature magnetic glassy phase, hysteresis cycles were measured as well (see \ref{ssec:DCmagnetization} and figure~\ref{fig:hyst}). The cycles show clear ``\textit{S}-shaped'' profiles for temperatures below $T_{\mathrm{irr}}$ for almost all the samples.
Measurements of \textit{ac susceptibility} were carried out on all the samples under test at three different frequencies (0.1, 1, and 10 kHz), with an ac excitation field $H_{ac}=1$~mT. No clear evidence of frequency dependence was found, except for the $x=0.21$ sample. Figure~\ref{fig:ACmagn} shows the low-temperature ac susceptibility data $\chi_{\mathrm{ac}}(T)$. At the lowest frequency, the temperature where $\chi_{\mathrm{ac}}$ shows a maximum is generally defined as the freezing temperature, $T_{f}$ \cite{MYDOSH}. We remark that this is similar to the $T_{\mathrm{irr}}$ of the dc measurements. As the frequency of ac measurements increases, we observe a continuous increase in $T_{f}$.
\subsection{\label{ssec:musr_exp}Muon-spin spectroscopy}
\begin{figure}
\centering
\includegraphics[width=0.375\textwidth]{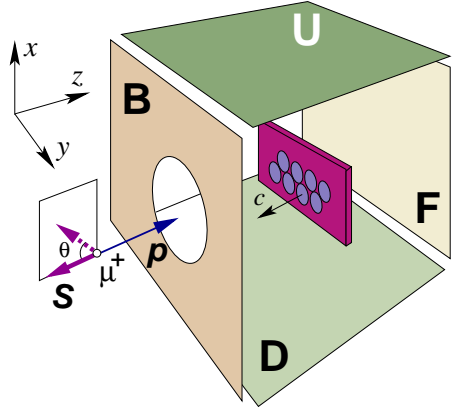} %geom_musr.eps
\caption{\label{fig:geom}(Colour online) Geometry of the ZF-$\mu$SR experiment showing the detector positions, the sample orientation, as well as the directions of muon momentum and the initial muon-spin polarization (normal and rotated).}
\end{figure}

Muon-spin spectroscopy ($\mu$SR) experiments were carried out at the GPS spectrometer ($\pi$M3 beam line) of the Paul Scherrer Institut, Villigen, Switzerland. A set of five single crystals was measured both in transverse-(TF) and in zero-field (ZF) configurations.
The technical details regarding the sample holder and the measurements are reported in \ref{ssec:holder}. A schematic drawing of the spectrometer, including the sample orientation, the beam direction and the detector positions is shown in figure~\ref{fig:geom}.
TF-$\mu$SR measurements were carried out by applying a (small) magnetic field perpendicular to the muon momentum $\boldsymbol{p}_{\mu}$. This configuration provides a straightforward way for estimating the paramagnetic volume fraction from the loss of spin polarization in the complementary magnetic part (see below).
ZF-$\mu$SR measurements were carried out using both a rotated and a nonrotated muon-spin polarization. In both configurations the samples were oriented so as to have the direction of the incident muon beam orthogonal to the sample surface and, therefore, parallel to the $c$ axis of \fetese\ crystals (see figure~\ref{fig:geom}).
\subsubsection{Transverse-field $\mu$SR}
All the selected crystals were investigated in a transverse field of 5 mT at two different temperatures: well below and above the respective N\'eel temperatures, $T_{\mathrm{N}}$. The latter were determined by dc magnetization measurements and confirmed also by ZF-$\mu$SR (see next paragraph).
In an applied transverse field, the coherent precession of muon spins in a nonmagnetic environment gives rise to a constant oscillation amplitude. On the other hand, muons stopping in magnetically ordered parts of the sample sense a wider distribution of fields, with a quick loss of coherence and consequent reduction (or even ``disappearance'') of the relevant asymmetry signal.
Due to the random character of muon implantation, the amplitudes of the different types of precession signals are proportional to the volume fraction of the respective environments.
Hence, by evaluating the ratio of the transverse asymmetries measured at $T \ll T_{\mathrm{N}}$ and at $T \gg T_{\mathrm{N}}$ we could determine the paramagnetic volume fraction, $V_{\mathrm{para}}/V_{\mathrm{tot}}$, as reported in table~\ref{tab:MagnProp}.
\begin{table*}[hbt]
\centering
\caption{\label{tab:MagnProp} Magnetic properties of various \fetese\ single crystals as determined from $\mu$SR and ac/dc magnetometry measurements.}
\begin{indented}
\lineup
\item[]\begin{tabular}{@{}*{7}{l}}
\br   
$x$(Se)&\textit{y}(Fe) &$T_{\mathrm{N}}$~(K)&$\Delta T_{\mathrm{N}}$~(K) &$V_{\mathrm{para}}/V_{\mathrm{tot}}$~(\%)&$T_{c}^{\mathrm{onset}}$~(K)&$f(t)$~(Eq.~\ref{eq:spin_prec})\cr
\mr
0.0  &0.087 &$67.6\0\pm 0.3$&$0.42\0\pm 0.09$&$33.3\0\pm 0.7$& ---           & $J_0$\cr
0.07 &0.05  &$43.6\0\pm 0.3$&$2.9\0\0\pm 0.4$&$17.8\0\pm 0.4$&$11.1\0\pm 0.5$& $J_0$\cr
0.21 &0.049 &$23.1\0\pm 0.3$&$3.1\0\0\pm 0.4$&$33.0\0\pm 0.6$&$12.1\0\pm 0.5$& $J_0$\cr
0.27 &0.053 &$17.6\0\pm 0.6$&$6.4\0\0\pm 0.9$&$32.6\0\pm 1.0$&$13.1\0\pm 0.5$& 1    \cr
\br
\end{tabular}
\end{indented}
\end{table*}
\begin{figure}
\centering
\includegraphics[width=0.5\textwidth]{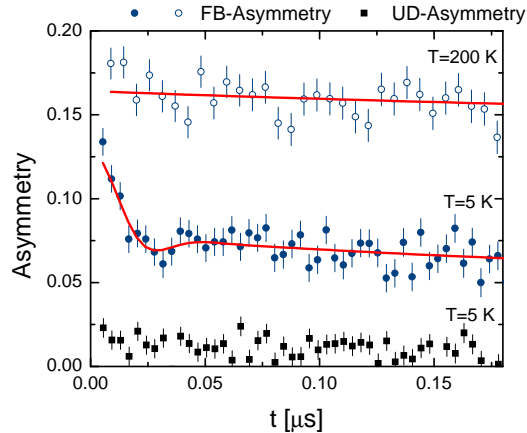} %asym10_NSR.eps
\caption{\label{fig:asym10_NSR}(Colour online) Low-temperature ZF-$\mu$SR
asymmetry as detected in the $x=0.07$ case by the F-B and U-D detectors in a configuration where the muon spin and momentum are collinear (and perpendicular to the iron planes).}
\end{figure}
\subsubsection{Zero-field $\mu$SR}
ZF measurements were carried out using either a rotated or a nonrotated muon-spin configuration, depending on the system availability. Typical asymmetry spectra for the $x=0.07$ case, collected at two selected temperatures in the forward-backward (F-B) detectors with a nonrotated muon spin, are shown in figure~\ref{fig:asym10_NSR}.
One can distinguish two different temperature regimes: in the low-temperature regime ($T < T_{\mathrm{N}}=43.6$ K) clear oscillations with a rather high damping rate are observed, demonstrating that the sample has made a transition to a magnetically ordered state characterized by long-range order. Here the oscillating signal is superimposed to a slowly decaying exponential relaxation, observable only at longer times (not shown). 
Above $T_{\mathrm{N}}$, instead, the amplitude of the oscillating signal goes abruptly to zero, leaving only two slow but distinct exponential relaxations. 
To follow the above evolution across the whole temperature range, all data sets were fitted to the function:
\begin{equation}
\label{eq:spin_prec}
A(t)= a_{T} \cdot f\left(B_{\mu},t\right) \cdot \exp(-\sigma^{2}_{T}t^2/2) + a_{L} \cdot \exp(-\lambda t) +\, a_{\mathrm{bg}}.
\end{equation}
Here the transverse component, characterized by an amplitude $a_{T}$ and a damping rate $\sigma_{T}$, is an oscillating function $f(B_\mu,t)=J_0(\gamma_{\mu} B_\mu t)$, with $J_0$ the zeroth-order Bessel function, $B_\mu$ the magnitude of the internal magnetic field, and $\gamma_{\mu} = 2\pi \times 135.5$ MHz/T the muon's gyromagnetic ratio. The longitudinal component was fitted to a relaxing Lorentzian term, with $a_{L}$ and $\lambda \sim 0.1$ $\mu$s$^{-1}$, the initial amplitude and relaxation rate, respectively, while $a_{\mathrm{bg}}$ is a temperature-independent relaxing background term that could be ascribed to a non negligible volume fraction that remains paramagnetic even at low temperatures (as inferred also from TF-$\mu$SR  data --- see previous section and table~\ref{tab:MagnProp}). 
It is worth noting that the highly damped oscillation found at short times below $T_{\mathrm{N}}$ in the $x=0$ sample could be fitted satisfactorily using either a cosine term or a Bessel function. Despite the long data-acquisition times, the collected statistics was still not high enough to clearly distinguish between these two cases. Consequently, although we can safely state that the system shows a long-range magnetic order, not much can be said about the commensurability of the order \cite{Khasanov2009}.
By taking into account the whole $\mu$SR data set and the theoretical predictions on the muon implantation sites (\textit{vide infra}), for the $x = 0$ case (pure FeTe) we privilege the fit results where the oscillating term was fitted with a zeroth-order Bessel function. This represents a coherent choice, in agreement with the results for the $x=0.07$ and $x=0.21$ samples, where the best fits were obtained with a $J_0$ term, too. As for the sample with $x=0.27$, here the damping was too strong and no oscillations could be detected at all [$f(t)=1$].
\begin{figure}[t]
\centering
\includegraphics[width=0.45\textwidth]{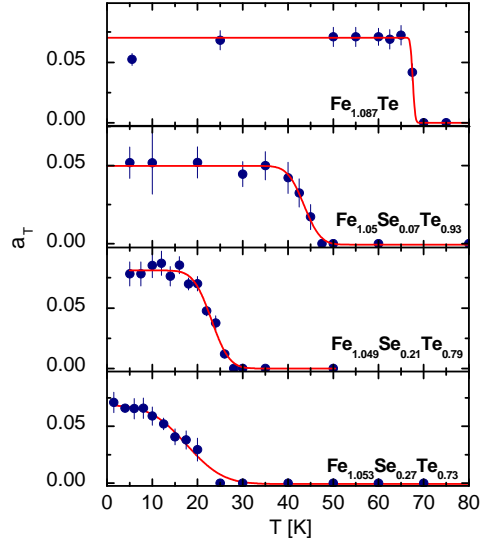} %Tneel.eps
\caption{\label{fig:Tneel}(Colour online) Temperature dependence of $a_T$, the amplitude of ZF-$\mu$SR transverse component, as measured in different \fetese\ single crystals. In each case the value of $a_T$ is proportional to the magnetically ordered volume fraction (maximal at $T=0$). All the data has been collected by the U-D detectors, except for the $x=0.07$ case, where no spin rotator could be used and the data refer to the forward-backward (F-B) detector pair. Solid lines represent numerical fits with an erf function (see text for details).}
\end{figure}
In figure~\ref{fig:Tneel} we summarize the temperature dependence of the %%%initial 
amplitudes of the transverse component $a_{T}$.
\begin{figure}[t]
\centering
\includegraphics[width=0.36\textwidth]{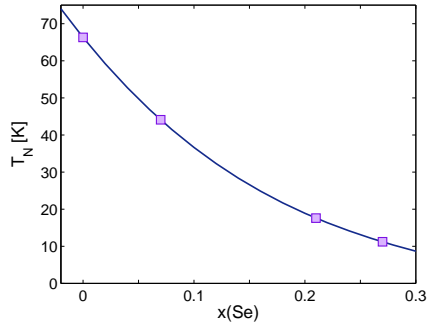} %Tn_summary.eps
\caption{\label{fig:Tn_summary}(Colour online) $T_{\mathrm{N}}$ versus Se content for all the tested samples, as determined by fitting the temperature dependence of the ZF-$\mu$SR transverse component $a_{T}$ with an erf function. The line is a guide for the eyes.}
\end{figure}
Since a transverse component in ZF data can arise only in presence of a magnetically ordered phase, by fitting the temperature-dependent amplitude $a_{T}$ with a complementary erf function, $1-\mathrm{erf}[(T-T_{\mathrm{N}})/(\sqrt{2}\Delta T_{\mathrm{N}})]$, 
we could determine the N\'eel temperature, $T_{\mathrm{N}}$, and the transition width, $\Delta T_{\mathrm{N}}$, of the ordered phase.
The resulting values for each sample are reported in table~\ref{tab:MagnProp} and in figure~\ref{fig:Tn_summary}. We note that $T_{\mathrm{N}}$ decreases progressively as the Se content increases.
\begin{figure}
\centering
\includegraphics[width=0.48\textwidth]{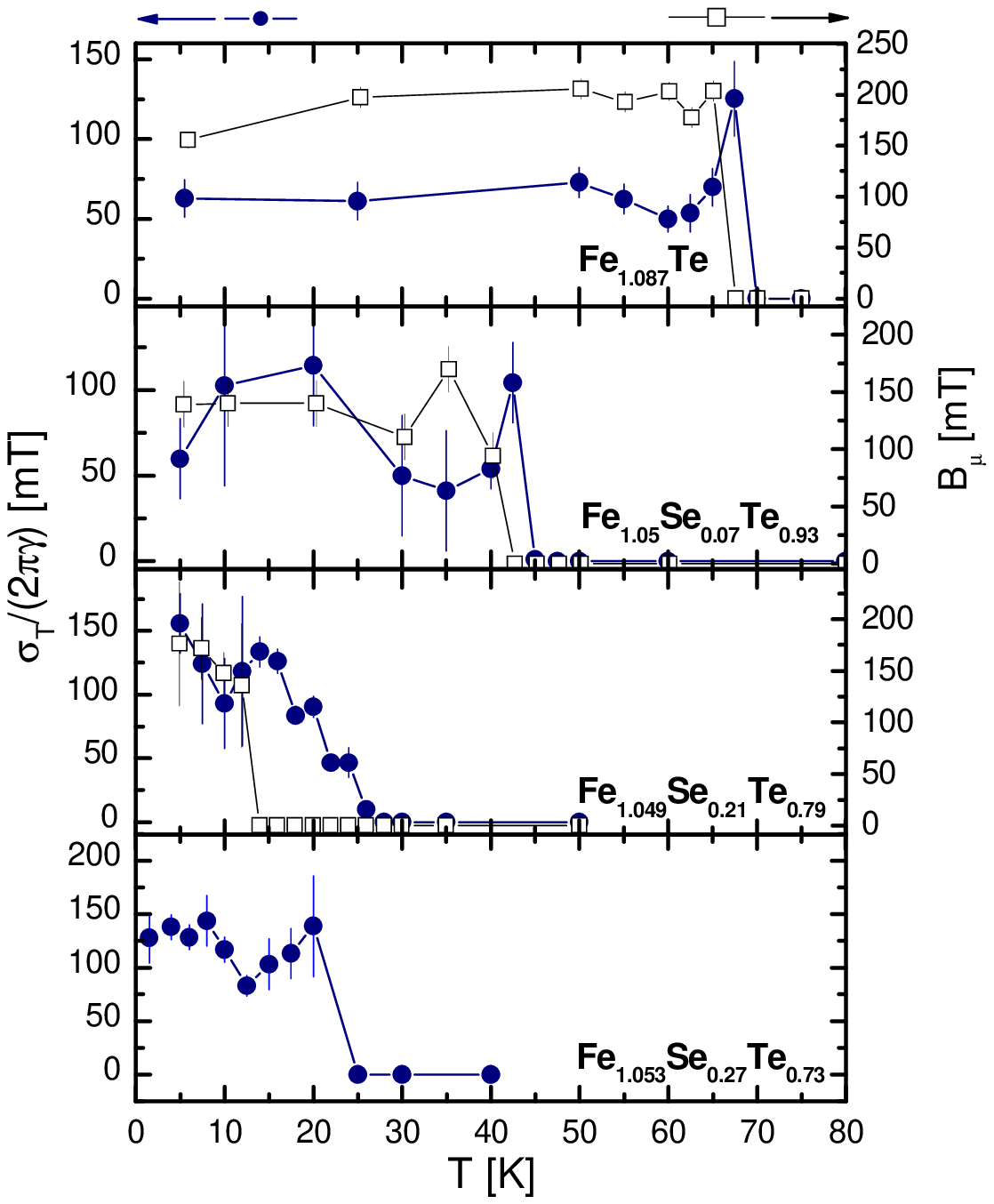}  %Bint.eps
\caption{\label{fig:Bint}(Colour online) Temperature dependence of the internal magnetic field $B_\mu$ (\opensquare) and of the transverse relaxation Gaussian width, $\sigma/\gamma_{\mu}$ (\fullcircle), for the investigated \fetese\ single crystals.}
\end{figure}
Figure~\ref{fig:Bint} shows the temperature dependence of the internal magnetic field $B_\mu$, as probed by the implanted muons.
The best fits at low temperature yield a mean-field value comprised between 100 and 200 mT for samples with $x=0$, 0.07, and 0.21 (right scale in figure~\ref{fig:Bint}). In case of $x=0.27$ the oscillations disappear and, in absence of $B_{\mu}$, the width of the Gaussian relaxation, $\sigma/\gamma_{\mu} = (\overline{B_\mu^2} - \overline{B_\mu}^2)^{1/2}$ \cite{Kadono2004,Shiroka2011}, can only provide an upper limit to the distribution of the internal fields at the muon site. In our case, at the lowest measured temperatures, these field values lie in the range 50 to 150 mT (left scale in figure~\ref{fig:Bint}).
\subsubsection{\label{sec:field_direction}Direction of the internal magnetic field}
In order to study the direction of the internal magnetic field $B_\mu$
in the differently doped samples we take advantage of the ZF-$\mu$SR as a non-perturbing technique (no applied external field).
To simplify the interpretation of the results we consider here
the case when the initial muon-spin polarization is antiparallel to the beam direction and systematically compare the data collected by the F-B with those of the U-D detectors (see figure~\ref{fig:geom}). We recall that each investigated sample consists of a mosaic of single crystals from the same batch, positioned so that their $ab$ surfaces coincide with
the $xy$ plane (i.e.\ orthogonal to the muon spin), as shown schematically in figure~\ref{fig:geom}. Therefore, the initial spin polarization is always parallel to the crystalline $c$ axis, while the $a$ and $b$ axes of the mosaic crystals are randomly oriented in the $xy$ plane. As an illustrative example, figure~\ref{fig:asym10_NSR} compares the muon-spin asymmetry measured in the up-down vs.\ forward-backward detector pairs for the $x=0.07$ case at $T = 5$~K. Similar results were obtained also for other samples and/or different temperatures (not shown). We note that the asymmetry measured by the U-D detector pair (i.e.\ along the $x$ direction) is practically zero.\footnote{Supposing $B_{\mu}$ to be in the sample's $ab$-plane, a small $c$-axis misalignment 
($\theta_{c} < 10^{\circ}$) could account for a tiny nonzero asymmetry in 
the U-D detectors: $a_{\mathrm{UD}} \propto a_{\mathrm{FB}} \cdot \tan \theta_{c} \simeq 0.02$ independent of time/temperature.} At the same time the signal detected by the F-B pair (i.e.\ along the $z$ direction) not only accounts for the total measured asymmetry, but its value $A_{\mathrm{tot}}(t)$ changes with time, reflecting a magnetically ordered state. These features indicates that the internal magnetic field $B_{\mu}$, as probed by the implanted muons, can be exclusively either in the $ab$ plane or orthogonal to it (i.e.\ not in some arbitrary out-of-plane direction). Now, if the internal magnetic field $B_\mu$ were parallel to the crystal $c$-axis, we would expect the asymmetry measured by the U-D and F-B detector pairs to be equal to zero and to a constant value, respectively. Since this is not the case, we conclude that the direction of the internal field lies necessarily in the laboratory $xy$ plane, i.e.\ in the sample's $ab$ plane. 
\section{\label{sec:discussion}Discussion}
\begin{figure}
\centering
\includegraphics[width=0.4\textwidth]{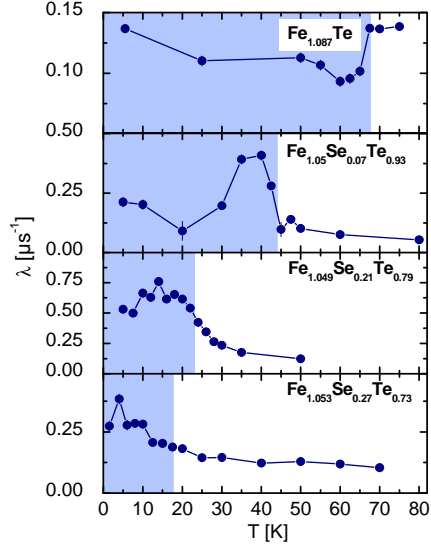}  %lambda1.eps
\caption{\label{fig:lambda1}(Colour online) Temperature dependence of the muon-spin longitudinal relaxation rate $\lambda$ for the investigated \fetese\ single crystals. The light blue areas indicate the magnetic phase.}
\end{figure}

\begin{figure}
\centering
\includegraphics[width=0.45\textwidth]{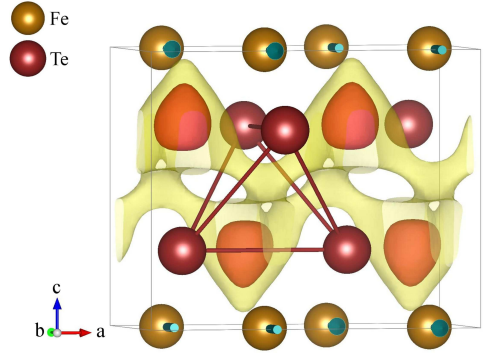}  %isosurfaces.eps %scale=0.14
\caption{\label{fig:muon_isosurfaces} Isopotential surfaces $V(r)$ for sites A and B calculated at the respective $V^{A,B}(r) = V^{A,B}_{min}+E^{A,B}_{0}$ 
(see text). The disjointed inner surfaces (red) refer to A-sites, while the interconnected sheets (yellow) to the B-sites.}
\end{figure}

We discuss now the experimental results presented so far with the aim of obtaining a unified picture for the entire \fetese\ series.\\
\noindent
\textit{DC magnetization}: as a general feature, the investigated samples display a complex magnetic behaviour. First, all the samples with $x>0$ are superconductors, whose $T_c$ and shielding fraction scale with $x$. Since the normalized SC shielding is always far away from the ideal value $-1$, even for the sample with the highest Se content, this suggests that superconductivity has not a bulk character. Second, $\chi(T)$ curves taken at 200 mT (large enough to destroy a weak superconductivity) exhibit a pronounced hysteresis, with irreversibility temperatures 
$T_{\mathrm{irr}}$ that decrease both with increasing Se content (instead of being independent of it, as previously reported, e.g., in Ref.~\cite{TranquadaJPSJ}) and with the applied field (not shown). In particular, the field dependence of $T_{\mathrm{irr}}$ is generally ascribed to the presence of a spin glass phase \cite{MYDOSH}. Such a hysteretic behaviour most likely originates from the competition between the antiferromagnetic \textit{intra}-cluster and the ferromagnetic \textit{inter}-cluster interactions. It suggests the lack of a true long-range magnetic order \cite{Fiorani1984,Mukherjee1998,Maignan1998,Dusan2000,Bai2005}, as confirmed also by our ZF-$\mu$SR experimental results (\textit{vide infra}).
Further insight on the glassy magnetic behaviour is obtained from the isothermal magnetization curves taken below and above the respective antiferromagnetic transitions (see figure~\ref{fig:hyst}). The low temperature data show two noteworthy features: \textit{i}) for $x\mathrm{(Se)}>0$ the first magnetization curve is compatible with the presence of a small superconducting phase. \textit{ii}) The hysteresis cycles consist of a complicated superposition of weak superconductivity (for $x>0$) and a glassy magnetic state, that give rise to characteristic ``\textit{S}-shaped'' $m(H)$ profiles \cite{MYDOSH,hyst_1,hyst_2}. This feature becomes more prominent at a higher Se content, whereas it seems hidden in the $x=0.07$ sample. The close lying superconducting and irreversibility temperatures make it very difficult to distinguish between these two coexisting effects.\\
\noindent
\textit{AC susceptibility}: only the ac susceptibility data of the $x=0.21$ sample display a clear frequency-dependent maximum at low temperature. Figure~\ref{fig:ACmagn} shows the temperature dependence of $\chi_{\mathrm{ac}}$, the ac magnetic susceptibility. The maximum of $\chi_{\mathrm{ac}}$, observed at freezing temperatures $T_f = 14$--15~K for frequencies $\nu =0.1$--10~kHz, represents the point where the system's relaxation time $\tau$ equals the characteristic time scale of the ac measurement, $1/\nu$. The ac magnetic response becomes frequency-independent only above that maximum, a behaviour peculiar of spin glasses \cite{MYDOSH}.
The $T_{f}$ value determined from the $\chi_{\mathrm{ac}}$ maximum is in very good agreement with %that 
the $T_{\mathrm{irr}}$ value extracted from the dc-magnetization FC-ZFC irreversibility curves at 200 mT (see figure~\ref{fig:DCmagn}).
We notice also that the peak position, $T_{f}$, shifts to higher temperatures with increasing frequency, yet another characteristic of spin glasses or superparamagnetic states. In fact, in these systems the relative freezing temperature variation, $\Delta T_f / T_f$, scales proportionally with the logarithmic frequency variation of the excitation frequency $\Delta \log(\nu)$ \cite{Liu2011,Tiwari2005,wang2006}:
\begin{equation}
\label{eq:freq_change}
\Delta T_{f} / T_{f}=\delta \cdot \Delta \log(\nu)
\end{equation}
The scaling parameter, $\delta$, usually lies between 0.005 and 0.06 in case of spin glasses, while it exceeds 0.1 in superparamagnets with noninteracting spin clusters. As shown in the inset of figure~\ref{fig:ACmagn}, the fit of experimental data provides $\delta=0.029(1)$, a value close to those found in other spin-glass/cluster-glass systems \cite{Liu2011,Song2008,BINDER,MYDOSH}. This rules out any possible superparamagnetic state and represents strong evidence about spin-glass behaviour in the \fetese\ family.\\
\noindent
\textit{Muon spectroscopy}: the low-temperature, short-time asymmetry data for $0\leq x \leq 0.2$ show a coherent muon-spin precession successfully fitted by a $J_0$ term. This feature is generally ascribed to the onset of a long-range AFM order, although somehow weakened by the incommensurate character of the magnetic structure \cite{Yaouanc2011} or, more simply, by the presence of magnetic clusters \cite{SAVICI,UEMURA}. 
In principle, our findings are in agreement with previous neutron diffraction \cite{Martinelli2010,PDai2010,PDai2009,Bao2009} and $\mu$SR measurement \cite{Khasanov2010,Khasanov2009} data, both suggesting a long-range AFM magnetic order. However, the more marked manifestations of a disordered magnetism in our case, lead us to associate the Bessel term of the fits with a generic medium-to-short range magnetism, as is the case of glassy magnetic phases. Indeed, the presence of large magnetic clusters has been invoked to explain the damped oscillations observed in certain conventional spin-glass systems \cite{SannaCLUSTER,CONERI}. This picture most likely applies whenever sufficiently big magnetic clusters can cause both coherent muon precessions, and give rise to a glassy magnetic hysteresis in the dc magnetization data. As the Se content is further increased beyond $x \sim 0.2$, there is a progressive loss of muon-spin polarization, reflected in an enhanced damping of the oscillating signal, which evidences a transition to a pure short-range magnetic order \cite{Yaouanc2011}.
Further hints towards the presence of a glassy magnetic phase can be inferred from the temperature behaviour of the longitudinal relaxation $\lambda_1$ (see figure~\ref{fig:lambda1}).
Here the broad peak that develops below $T_{\mathrm{N}}$ can be explained in terms of a progressive freezing of the magnetic correlations. In general, magnetic correlations with a glassy character were found in all the samples under test. Indeed, the typical features expected in a classical spin-glass phase (such as, e.g., the bifurcation in dc magnetization) are present in all of them, but they are most pronounced in the $x=0.21$ case, which fulfills at best the criteria defining the spin-glass state (see also Figs.~\ref{fig:ACmagn} and \ref{fig:hyst}). In this sense, by considering only dc magnetization data, our results fully agree with the phase diagram reported in Ref.~\cite{TranquadaJPSJ}. However, by considering also the additional data of ac susceptibility, muon spectroscopy and dc isothermal magnetization, we obtain a scenario where \fetese\ seems to display a generalized magnetic glassy phase, characterized by hysteretic magnetic correlations from $x=0$ up to $x=0.27$. 
As detailed in our final remarks, the difficulty in unambiguously defining the spin-glass state in \fetese\ is most likely due to the presence of local structural, chemical and electronic inhomogeneities, intrinsic to this class of materials \cite{Joseph2010,Iadecola2011,Hu2011}.
As described in Sec.~\ref{sec:field_direction}, our experimental data suggest that the internal magnetic field, as sensed by the precessing muons, lies in a plane parallel to the iron layers. To better appreciate the implications of this finding we tried to determine the muon-stopping site(s) by investigating an ideal stoichiometric FeTe crystal with the Density Functional Theory (DFT) formalism in the generalized gradient approximation (GGA) \cite{PBE}. We adopted the Full Potential Linearized Augmented Plane Waves (FP-LAPW) method, as implemented in the Elk code \cite{elk_code}. In particular, we considered a bicollinear magnetic order on a $2a \times a \times c$ cell with $P4/nmm$ symmetry \cite{Martinelli2010} and a 
$R_{\mathrm{min}}^{MT}\times \mathrm{max}(|\mathbf{k}|) = 8$ is chosen
for the expansion of the wave functions in the interstitial region
($R_{\mathrm{min}}^{MT}$ is the smallest muffin-tin radius in the unit
cell and $\mathrm{max}(|\mathbf{k}|)$ the largest wave number of the
basis set). The resulting magnetic moment per iron atom was 2.24 $\mu_{\mathrm{B}}$, in 
reasonable agreement with 2.54 $\mu_{\mathrm{B}}$, the experimental value obtained via 
neutron powder diffraction \cite{Martinelli2010}.
\begin{figure}
\centering
\includegraphics[width=0.45\textwidth]{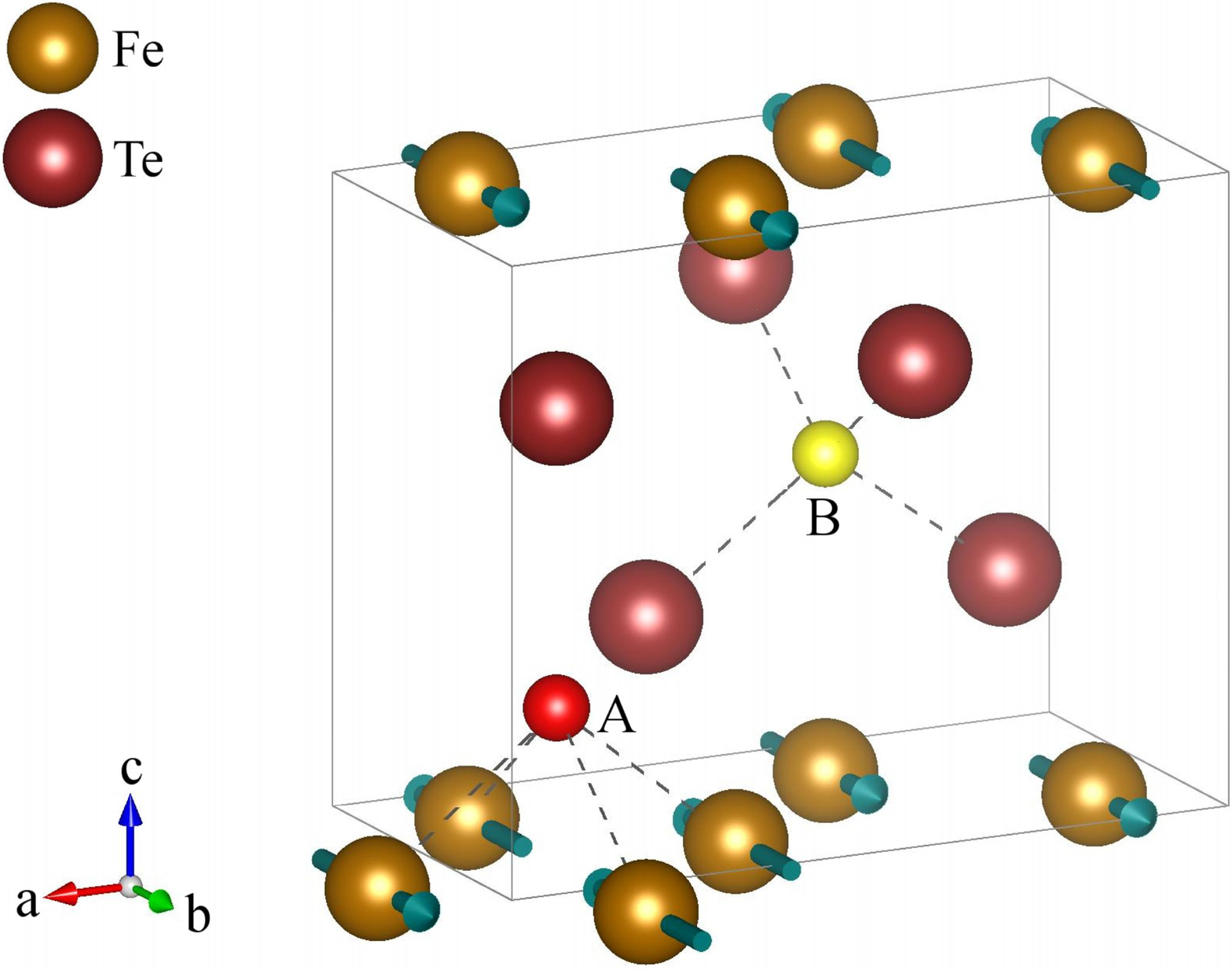}  %sites.eps %scale=0.14
\caption{\label{fig:muon_sites}Possible muon stopping sites within the tetragonal unit cell (high-temperature paramagnetic phase), as obtained from Coulomb potential minima. 
The positions of Fe and Te atoms were taken from Ref.~\cite{Martinelli2010}. The arrows represent Fe-ion magnetic moments in the ordered phase. Upon considering the $\mu^{+}$ zero-point motion, only sites of type A are found to be stable.}
\end{figure}
To determine the muon-stopping sites we considered the minima of the sign-reversed electrostatic potential, as obtained from a self-consistent DFT calculation. The positively charged muons will thermalize in the interstitial positions corresponding to the potential minima. Two inequivalent minima were found, whose positions are shown in figure~\ref{fig:muon_sites}, with the respective coordinates reported in table~\ref{tab:muonpos}. The absolute minimum (site A) replicates the one determined in 
Ref.~\cite{Bendele2012} for FeSe, while the secondary minimum (site B) coincides with the centre of the tetrahedron formed by Te atoms. The latter site turned out to be unstable against the zero-point motion of the muon. This was accounted for by modelling the potential minimum with an anisotropic harmonic well (AHW)  
$V(r)=\frac{1}{2}m(\omega^{2}_{x}x^{2}+\omega^{2}_{y}y^{2}+\omega^{2}_{z}z^{2})$ and evaluating the zero-point energy $E_{0}$ of the implanted muon.
The resulting isosurfaces for $V^{A,B}(r) = V^{A,B}_{\mathrm{min}}+E^{A,B}_{0}$ are presented in figure~\ref{fig:muon_isosurfaces}. While A-site minima remain on disconnected isosurfaces, those for B sites form an interconnected network that rules out the presence of muon bound states. Therefore, also those muons initially implanted in sites of type B will migrate towards the nearest A site. Hence, sites of type A represent the only possible muon implantation site in the FeTe crystal. We recall that sites A almost coincide with the positions of Fe(II) but, since their iron occupancy is lower than 10\% for all the considered $y$ compositions (see also table~\ref{tab:MagnProp} and Ref.~\cite{giannini2010}), they are widely available for hosting the implanted muons. 
\begin{table}
\caption{\label{tab:muonpos}Properties of the muon-stopping sites. The reported positions correspond to the minima of the potential $V(r)$. The respective Wyckoff sites are A $2c$ and B $2b$. $E_{0}$ represents the zero-point energy (see text) and WFC is the wave function centre.}
\begin{indented}
\lineup
\item[]\begin{tabular}{@{}*{5}{l}}
\br 
Site & Position & $z_{\mu^{+}}$ WFC & $V-V_{A}$ (eV) & $E_{0}$ (eV)\cr
\mr
A & ($\nicefrac{1}{4}$, $\nicefrac{1}{4}$, 0.245) & 0.27 & 0 & 0.36 \cr
B & ($\nicefrac{3}{4}$, $\nicefrac{1}{4}$, 0.500) & 0.50 & 0.9 & 0.39 \cr
\br
\end{tabular}
\end{indented}
\end{table}
Note that the electrostatic potential has an anharmonic component along the $c$ direction. Therefore, the equilibrium position of the muons may not correspond to the potential minimum. Indeed, considering the anharmonicity, we find that the correct value for the site A is ($\nicefrac{1}{4}$,$\nicefrac{1}{4}$, 0.27), as obtained from the centre of the ground-state wavefunction.
Taking this into account, we can evaluate the local field probed by the implanted muons, $\vec{B}_{\vec{\mu}}$, which consists mostly of two contributions, $\vec{B}_{\vec{d}}$ and $\vec{B}_{\vec{c}}$, the so-called dipolar and Fermi-contact fields, respectively \cite{Yaouanc2011}. If we assume that the behaviour of \fetese\ series does not differ much from that of pnictides, then the Fermi contact contribution is expected to be negligibly small \cite{Maeter2009}. The dominant $\vec{B}_{\vec{d}}$ term at the muon-stopping site was calculated by means of the probability density obtained from the AHW ground-state wavefuction. By considering the ordered configuration of Fe(I) magnetic moments (each carrying 2.54 $\mu_{\mathrm{B}}$, as from neutron powder diffraction on Fe$_{1+y}$Te polycrystalline samples) \cite{Martinelli2010}, the expected dipolar field at the muon implantation sites is $\sim 238$~mT. The latter result is in good agreement with $197 \pm 11$~mT, the value obtained from ZF-$\mu$SR data. The residual discrepancy can be attributed to a possible small contribution arising from the Fermi-contact term, that was neglected in a first instance.
We recall here that the excess of iron (present in all samples) is strongly magnetic \cite{Zhang2009} and may play an important role in broadening the local field distribution. Moreover, as shown in figure~\ref{fig:Bint}, the low-temperature field value at a muon site, $B_\mu$, decreases slightly with increasing Se content. 
Provided the hyperfine coupling tensor does not depend (or depends only weakly) on Se content, this would imply that the magnetic moment of the iron ions does the same, in apparent contrast with neutron powder diffraction results \cite{Martinelli2010}.\\ \\
Finally, we recall that EXAFS\cite{Joseph2010,Iadecola2011}, STEM and EELS \cite{Hu2011} results, previously mentioned in the introduction, bring strong and clear evidence about local structural inhomogeneities, nanometer-scale phase separation and chemical inhomogeneities on \textit{crystallographically homogeneous} \fetese samples. These important findings let us unify our experimental results into a coherent picture.
Since X-ray diffraction measurements evidence that our samples are in a single phase \cite{giannini2010}, the above mentioned nanoscale phase separation scenario would imply: \textit{i}) the formation of magnetic clusters, hence justifying the glassy behaviour evidenced by both ac/dc magnetometry and muon spectroscopy data. This same glassy behaviour could potentially attenuate, or even mask, the true variation of the Fe magnetic moment with Se/Te substitution. \textit{ii}) The presence of a non negligible volume fraction that stays in the paramagnetic state also at low temperatures. In our case, this fraction represents ca.\ 20--30\% of the sample volume and is mostly Se-independent. \textit{iii}) The absence of any possible nanoscopic-like M-SC coexistence, rather common instead in the 1111 pnictide family \cite{Sanna2009,Sanna2010Rapid,Sanna2011}. The latter can be ruled out because none of the investigated SC samples shows bulk superconductivity. In particular, this seems to be a general feature of the iron chalcogenide family that confirms \textit{a posteriori} previous muon spectroscopy and neutron diffraction results on \fetese\ single crystals \cite{Khasanov2009,Khasanov2010}.

\section{Conclusion}
The low-temperature antiferromagnetic order in an extended series of \fetese\ single crystals was studied by means of ac/dc magnetometry and by muon-spin spectroscopy. Muons are most likely implanted close to a Fe(II) site, where they sense a local magnetic field $B_\mu \sim 200$~mT, lying in the $ab$ plane.
At low Se content $\mu$SR detects a medium/long range AF order that progressively weakens with Se substitution and, for $x > 0.2$, turns into a short-range order. Both ac and dc magnetometry data on the same samples show clear evidence about a magnetic glassy phase, which becomes more prominent for $x=0.21$, but vanishes at higher Se content. 
The magnetometry results, combined with the unusually highly-damped ZF-$\mu$SR oscillations, observed in the same temperature range where the glassy behaviour occurs, suggest that the system is most likely in a \textit{magnetic cluster-glass state}, probably induced by the  nanometer-scale phase separation and the chemical inhomogeneity. The last feature seems to be an intrinsic property of the \fetese\ family, as confirmed also by recent extensive studies on polycrystalline FeSe$_{0.5}$Te$_{0.5}$ \cite{Pal2012}.

\ack
This work was performed at the Swiss Muon Source S$\mu$S, Paul Scherrer Institut (PSI, Switzerland). The authors are grateful to A.~Amato for the instrumental support and helpful discussions. G.L.\ thanks A.\ Martinelli for fruitful discussions and suggestions. T.S.\ acknowledges support from the Schweizer Nationalfonds (SNF) and the NCCR program MaNEP. This work was partially supported by Grant No. PRIN2008XWLWF9. F.B. acknowledges support from CASPUR under the Standard HPC Grant 2012.

\section*{Appendix}
\appendix
\section{\label{ssec:DCmagnetization}Supplementary information on DC magnetization}
\begin{figure}
\centering
\includegraphics[width=0.50\textwidth]{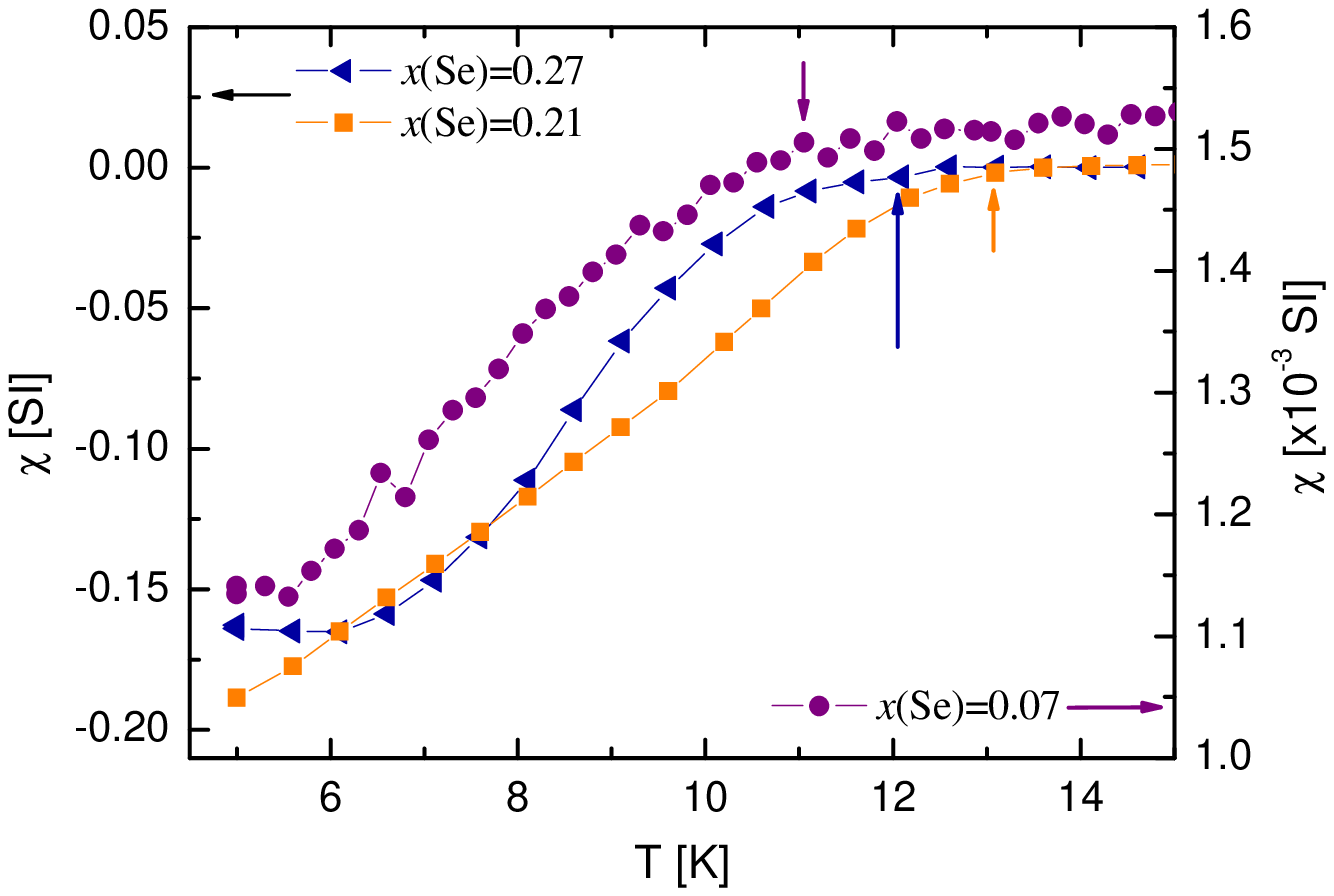} %supercond.eps
\caption{\label{fig:supercond} (Colour online) Shielding volume fraction for all the Se-substituted samples obtained from ZFC curve as measured by dc SQUID. The data were taken in an applied magnetic field of 0.5 mT (except for the $x=0.07$ sample, which was measured at 10 mT). The vertical arrows indicate the onset of superconductivity.}
\end{figure}

\begin{figure*}
\centering
\includegraphics[width=0.35\textwidth]{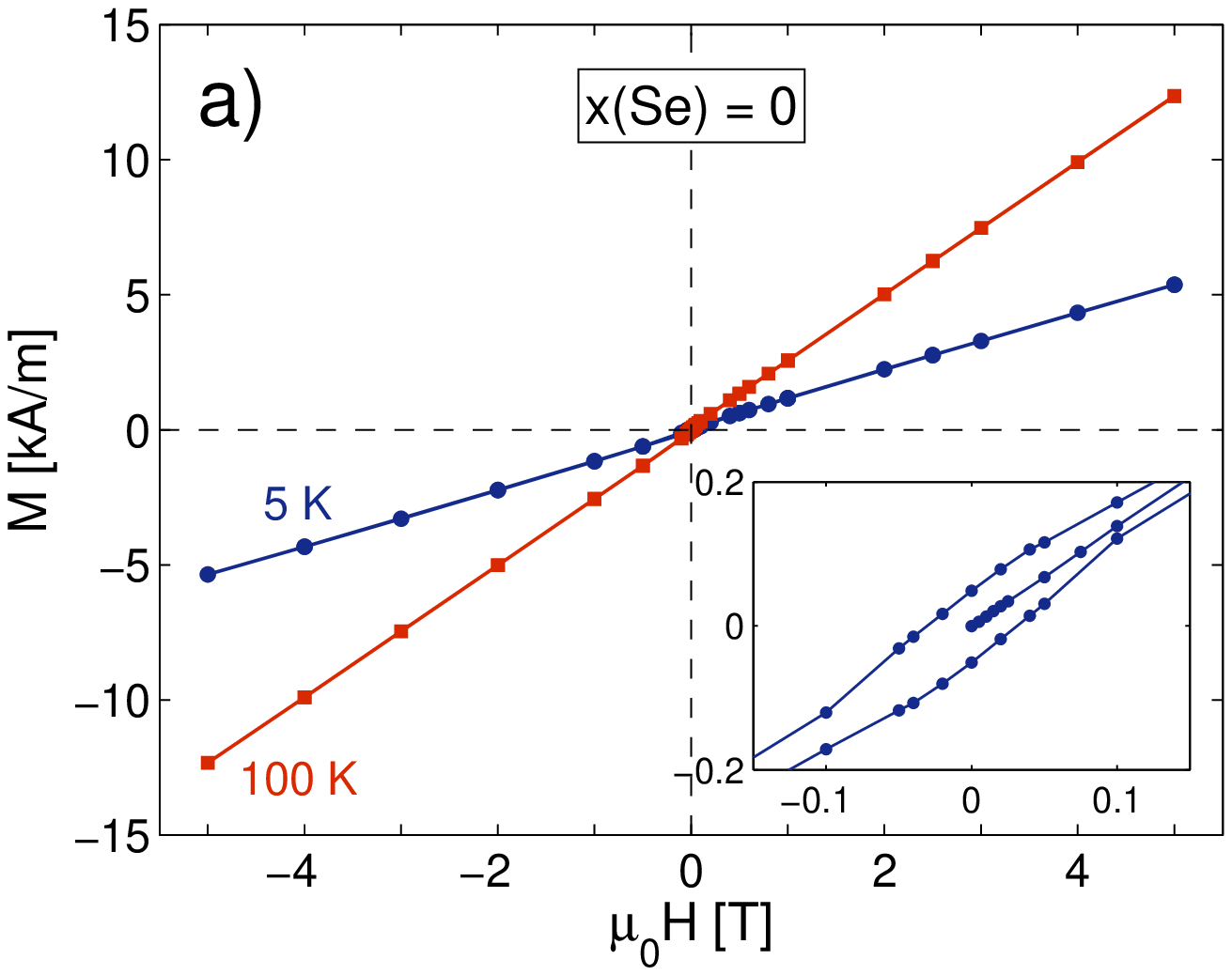}  %hyst_x000.eps
\hspace{5mm}
\includegraphics[width=0.35\textwidth]{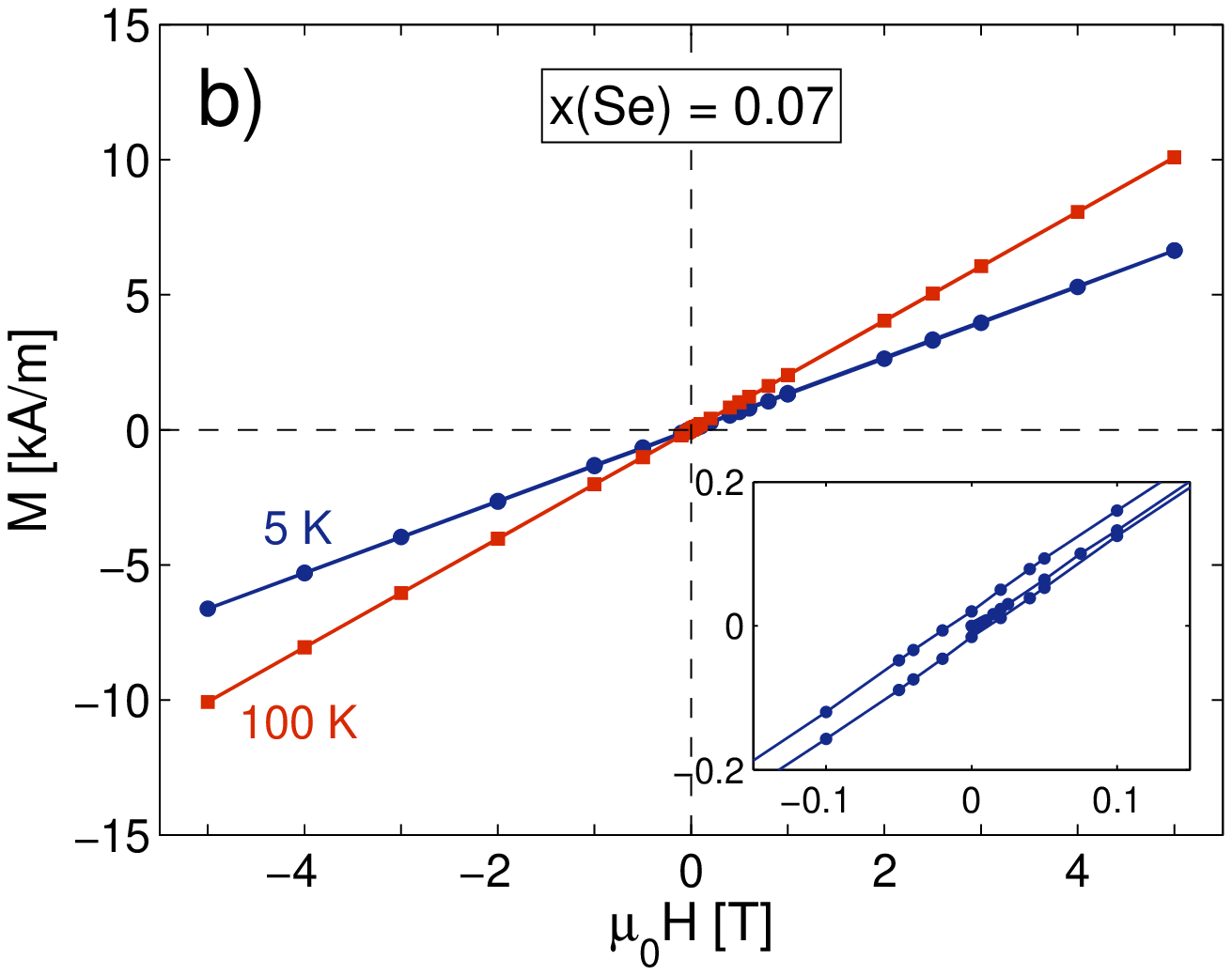}  %hyst_x007.eps
\vskip 5mm
\includegraphics[width=0.35\textwidth]{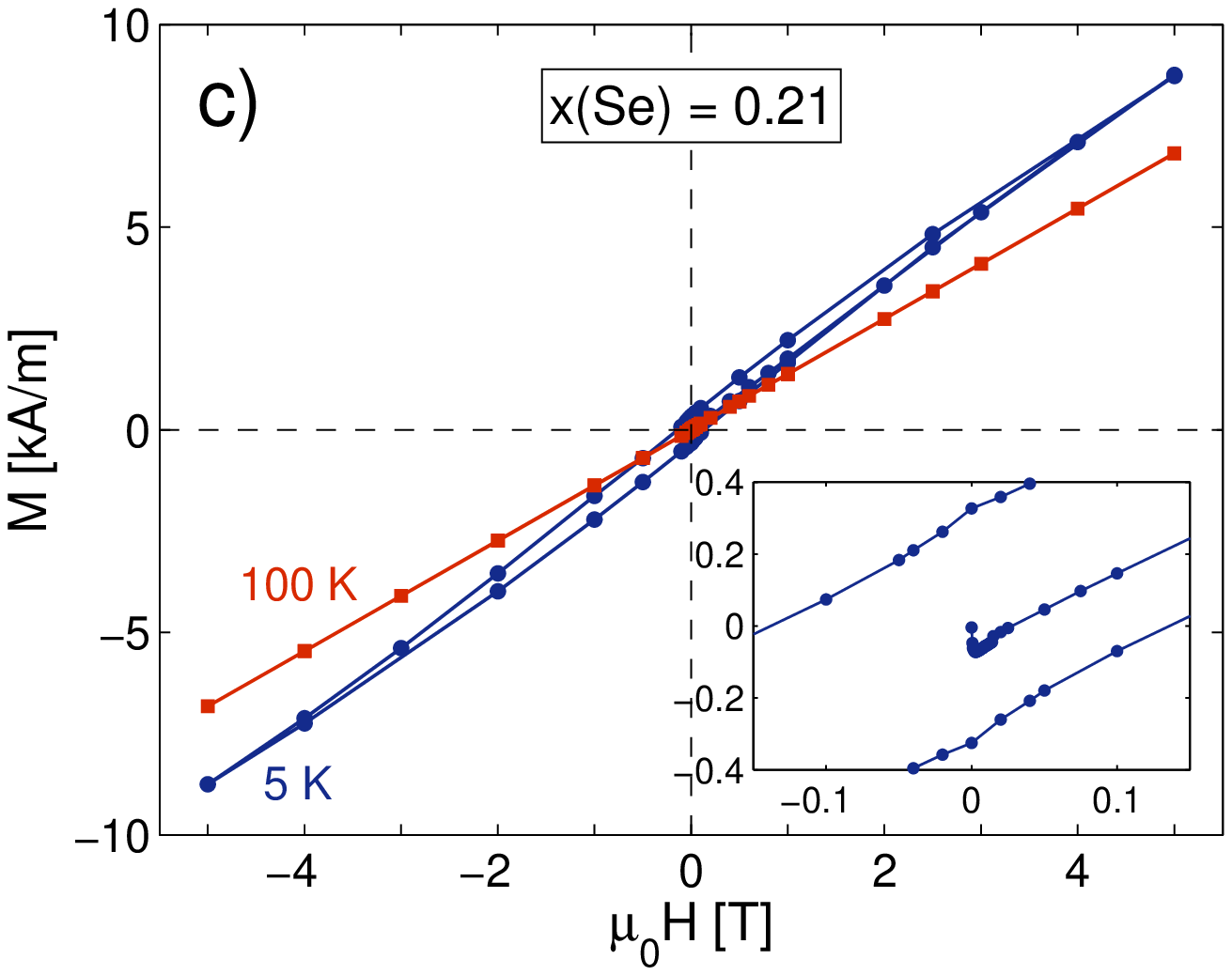}  %hyst_x021.eps
\hspace{5mm}
\includegraphics[width=0.35\textwidth]{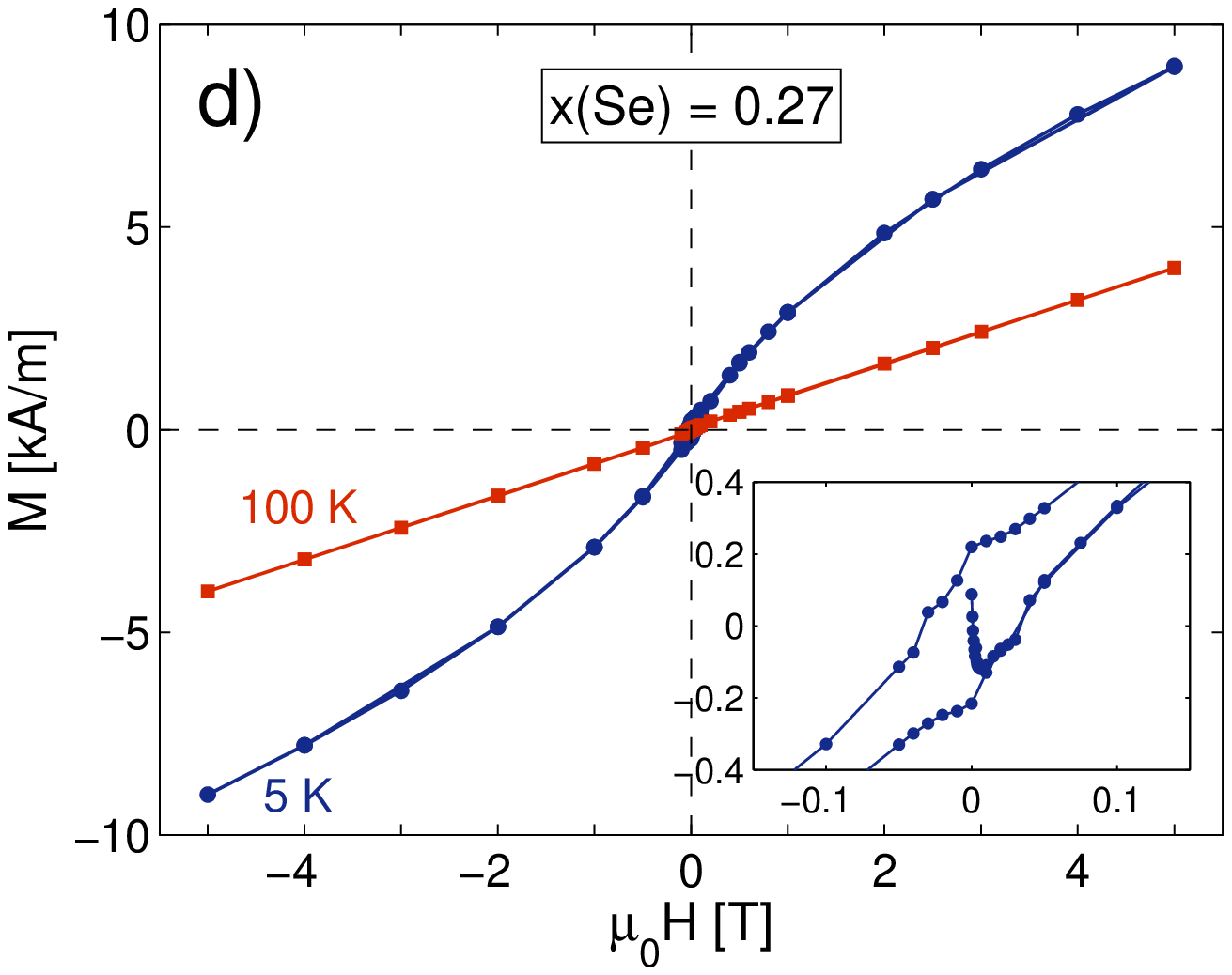}  %hyst_x027.eps
\caption{\label{fig:hyst} (Colour online) Hysteresis cycles at 5 and 100 K for all the \fetese\ samples under study. Insets show an expanded view of the magnetic field dependence of the first magnetization curve at 5 K.}
\end{figure*}

We present here supplementary material on dc magnetization measurements. It consists in:~\textit{a}) the low-field temperature dependence of the dc magnetization, reported in figure~\ref{fig:supercond}, and \textit{b}) the hysteresis cycles measured below and above $T_{\mathrm{N}}$, as determined by ZF-$\mu$SR spectroscopy, reported in figure~\ref{fig:hyst}.
\section{\label{ssec:holder}The $\mu$SR sample holder and measurement details}
The $\mu$SR measurement of single crystals with a typical size of $3 \times 4$ mm$^2$ and 0.1--0.3 mm thickness was made possible by using a standard ``mosaic technique''. Several crystals from the same batch were sticked close together using an aluminated mylar tape. The resulting mosaic was then centred on a standard aluminium fork sample holder suited for muon fly-past measurements. For all the investigated samples the effective sample area exposed to the beam was never below 25 mm$^2$. Taking into account the relatively large sample thickness ($\geq 0.25$ mm) and the average sample density, a 0.6 mm thick kapton mask was sufficient to ensure that the majority of the incoming muons were implanted into the sample volume. The further use of veto counting (which accounts for only those muons effectively stopped in the sample), contributed also to a significantly improved signal-to-noise ratio. In this configuration a total asymmetry in the range 0.15--0.22 could be obtained,
still low if compared with the normal asymmetry value expected at the GPS
instrument ($\sim 0.28$). The discrepancy could be ascribed to muonium (or muonated radical) formation in the kapton mask with a relatively high hyperfine coupling constant. Since the ensuing muon-spin depolarization reduces only the total signal asymmetry, but does not otherwise affect the measured signal, we could safely ignore the fraction of muons being stopped in the kapton frame.

\section*{References}

\providecommand{\newblock}{}

\end{document}